\documentclass[12pt]{article}
\usepackage{verbatim,graphicx,amssymb,cite}
\newcommand{\be}{\begin{equation}}
\newcommand{\ee}{\end{equation}}

\begin{document}
\setcounter{page}{0}
\begin{titlepage}
%\vspace*{-2.0cm}
\vspace*{0.1cm}
\begin{center}
{\Large \bf Neutrino nonstandard interactions in the supernova 
}\\
\vspace{1.0cm}

{\large
C. R. Das\footnote{E-mail: crdas@cftp.ist.utl.pt},
Jo\~{a}o Pulido\footnote{E-mail: pulido@cftp.ist.utl.pt}\\
\vspace{0.15cm}
{{\small \sl CENTRO DE F\'{I}SICA TE\'{O}RICA DE PART\'{I}CULAS (CFTP)\\
 Departamento de F\'\i sica, Instituto Superior T\'ecnico\\
Av. Rovisco Pais, P-1049-001 Lisboa, Portugal}\\
}}
\vspace{0.25cm}
\end{center}
\vglue 0.6truecm

\begin{abstract}
Neutrino non-standard interactions (NSI) were investigated earlier in the solar case and were shown to 
reduce the tensions between the data and the large mixing angle solution predictions. We extend the previous
framework to the supernova and evaluate the appearance probabilities for neutrinos and antineutrinos as a 
function of their energy after leaving the collapsing star with and without NSI. For normal hierarchy the
probability for electron neutrinos and antineutrinos at low energy ($E\lesssim 0.8-0.9 MeV$) is substantially
increased with respect to the non-NSI case and joins its value for inverse hierarchy which is constant 
with energy. Also for inverse hierarchy the NSI and non-NSI probabilities are the same for each neutrino and 
antineutrino species. Although detection in such a low energy range remains at present an experimental 
challenge, it will become a visible trace of NSI with normal hierarchy if they exist. On the other hand 
the neutrino decay probability into an antineutrino and a majoron, an effect previously shown to be induced 
by dense matter, is, as in the case of the sun, too small to be observed as a direct consequence of NSI.
\end{abstract}

\end{titlepage}

\section{Introduction}

Non-standard neutrino oscillations (NSI) \cite{Guzzo:1991hi,Roulet:1991sm,Grossman:1995wx,
Johnson:1999ci,Huber:2001zw,Huber:2002bi,Blennow:2008er,Biggio:2009nt,Wei:2010ww} have since long 
been applied to solar 
neutrinos \cite{Guzzo:1991hi,Roulet:1991sm,Pulido:1992nn} in an attempt to understand the origin of their 
apparent deficit. More recently interest in NSI has been revived \cite{Pulido:2010ht,Palazzo:2011vg}
%\footnote{For an extensive although 
%incomplete list of references on NSI see ref.\cite{Pulido:2010ht}.} 
with the purpose of solving the possible 
inconsistencies between the LMA solution to the solar neutrino problem \cite{Fogli:2008ig,Schwetz:2008er} and 
the data \cite{:2008zn,Smy,Aharmim:2009gd,Bellini:2010gn}. The most remarkable of these inconsistencies is the
absence of any experimental evidence for an upturn in the LMA survival probability in the intermediate energy 
range of solar neutrinos. It has been shown \cite{Pulido:2010ht} that NSI can lead to a flat probability in 
this sector and hence a flat electron energy spectrum as observed in both SuperKamiokande \cite{:2008zn} and SNO 
\cite{Aharmim:2009gd} experiments, with the interesting consequence of a possible neutrino decay into 
antineutrinos and majorons in dense matter. However, the predicted antineutrino flux is too small to be 
observed and the only expected signature of the effect is the flatness of the electron spectrum. Extension
to supernova neutrinos, which will be done in the present paper, will provide us further information on 
possible experimental NSI signatures. Our approach to NSI, previously developed in \cite{Pulido:2010ht} assumes 
extra contributions to the vertices $\nu_{\alpha} \nu_{\beta}$ and $\nu_{\alpha} e$ and differs 
therefore from the neutrino-neutrino interaction considered in \cite{Blennow:2008er}.
%For supernova 
%neutrinos, despite the much larger matter density, the antineutrino production probability also fails
%to be sizeable enough for an antineutrino flux from NSI to be observed, as will be seen in the present paper.

The supernova dynamics has been extensively studied and for details we may refer the reader to the review 
\cite{Bilenky:2002aw}. Here we will just highlight its relevant aspects for our purposes.
As is well known when a star of mass $M\gtrsim 8M_{\odot}$ has burned all its fuel, an onion like 
structure is formed with the lighter elements in the outer layers and the heavier ones inside. For massive 
enough stars $M\gtrsim 11M_{\odot}$ an innermost iron core is formed. Equilibrium of the core is disrupted 
by photodissociation of the heavier elements producing alpha particles and neutrons. The resulting free 
electrons are in turn captured by protons and nuclei, producing electron neutrinos which escape,
\be
e^{-}~p \rightarrow n~\nu_e~.
\label{neutron}
\ee
When the Coulomb pressure of the electrons becomes insufficient to sustain the core against its own gravity, 
collapse is initiated until the core becomes no longer transparent to neutrinos and reaches nuclear density
($\rho\simeq 3\times 10^{14}g~cm^{-3}$), a process which takes only about 10 ms \cite{Thompson:2002mw}. The 
neutronization process (\ref{neutron}) ceases, a state of hydrostatic equilibrium is reached with the core 
forming a proto-neutron star of radius around 10 km and a temperature of 35-40 MeV. %The layer beyond which 
%neutrinos can escape without scattering is called the "neutrino sphere". 
From this stage onwards only thermal neutrinos are emitted. They are produced from nucleon-nucleon bremsstrahlung
\begin{equation}
N~N\rightarrow N~N~\nu~\bar\nu
\end{equation}
$e^{+}e^{-}$ annihilation,
\begin{equation}
e^{+}e^{-} \rightarrow \nu~\bar\nu
\end{equation}
plasmon decay,
\begin{equation}
\gamma \rightarrow \nu~\bar\nu
\end{equation}
and electron nucleon bremsstrahlung
\begin{equation}
e^{-}~N \rightarrow e^{-}~N~\nu~\bar\nu.
\end{equation}
So according to core collapse supernova dynamics, in the thermal phase neutrino emission proceeds through
all flavour channels and is accompanied by antineutrinos, whereas in the initial neutronization phase only 
electron neutrinos are expected. 
%Hence the possible dense matter neutrino decay through NSI in the core can
%show up in a clear electron antineutrino signature in the neutronization phase. In fact, while antineutrinos 
%of all flavours are produced through dense matter decay, since supernova neutrino energy is of the order of
%(10-20) MeV, only $\bar\nu_e$'s can clearly be seen through the charged current process
%\be
%\bar\nu_e~p\rightarrow n~e^{+}~.
%\label{antinu}
%\ee
%Other antineutrinos ($\bar\nu_{x}$ for $x=\mu,\tau$) are detected through the neutral current which cannot 
%distinguish between them and leads to a much smaller event rate. 

The events observed in SN1987a were all interpreted as antineutrino ones \cite{Sato:1987rd} due not only 
to the dominance of the antineutrino cross section relative to the neutrino one but also possibly to the 
unavailability of a lower energy threshold, as shall be seen in the present paper. Whether or not any 
neutrinos reached the Earth is an open question. 
%The increased 
%sensitivity and availability of detectors nowadays will make it likely under similar circumstances to clearly 
%detect electron neutrinos from the supernova. 
In ref.\cite{Pulido:2010ht} it was shown that neutrino decay in dense matter into antineutrinos and a 
majoron is a necessary consequence of NSI. However its effect through the detection of antineutrinos in 
the neutronization phase where only $\nu_e$'s are produced is not possible, owing to the smallness of 
the antineutrino production probability through decay even in the supernova.
%Although neutrino decay into antineutrinos in dense matter is a necessary consequence of NSI, 
%its effect through the detection of antineutrinos in the neutronization phase is not possible, owing to
%the smallness of the antineutrino production probability through decay, as will be seen. 
In fact the high density range of the neutrino trajectory is not long enough to produce a visible antineutrino 
flux originated from this source. 
%The possible NSI signature will instead be the appearance of a significant 
%electron neutrino flux in the energy range 0.8-0.9 MeV and below, both in the neutronization phase and in the 
%thermal phase. Despite the increased sensitivity and availability of the detectors nowadays, we need, as in 
%the case of solar neutrinos, a decrease in the neutrino energy detection threshold in order to test this 
%interesting energy range.
 
%In this respect the NSI signature is similar 
%to the resonant spin flavour precession (RSFP) one \cite{Lim:1987tk,Akhmedov:1988uk} where a significant 
%amount of $\bar\nu_{e}'s$ is expected to be seen in the neutronization phase \cite{Akhmedov:2003fu}. 
%However while in RSFP the $\bar\nu_{e}$ energy is the same as the emitted $\nu_e$ one, in the NSI
%matter decay this energy is shared between the antineutrino and the majoron, so that a continuous
%spectrum arises with more $\bar\nu_{e}'s$ at the lowermost energies \cite{Pulido:2010ht}. 
%%However whereas $\bar\nu_{e}'s$ from RSFP are emitted with a single energy which is the initial 
%%$\nu_e$ one, for the NSI matter decay the initial $\nu_{e}$ energy is carried by the antineutrino and 
%%the majoron, so that a continuous energy spectrum is generated for the emitted $\bar\nu_{e}'s$ with  
%%more $\bar\nu_{e}'s$ at the lowermost energies \cite{Pulido:2010ht}. 
%Thus the signatures of the RSFP and NSI matter decay effects will be clearly distinguishable from each other.

In this paper we extend for supernova neutrinos the NSI framework that was
introduced in the solar neutrino case \cite{Pulido:2010ht} where it was shown to improve the fits to experiment.
In section 2 we review its general features. We recall that a flat SuperKamiokande and SNO electron spectrum  
necessarily require imaginary diagonal entries in the NSI Hamiltonian. Their real parts, along with the 
off diagonal entries whether real or imaginary, may be arbitrary in the sense that they do not induce any 
change in the standard LMA probability. As a consequence neutrino decay in dense matter into an antineutrino 
and a majoron arises. Section 3
%In this paper we examine the possible consequences of NSI neutrino decay in dense matter for the 
%supernova neutrino signal. In section 2 we review the general features of the NSI framework developed
%in ref.\cite{Pulido:2010ht}. We recall that the diagonal entries of the NSI Hamiltonian need to be
%imaginary. Their real parts, along with the off diagonal entries whether real or imaginary, may be 
%arbitrary in the sense that they do not induce any change in the standard LMA probability. Section 3
is the main part of our work. Here we extend NSI to the interactions of the supernova neutrinos. 
We evaluate the survival and conversion probabilities with and without NSI as well as the probability 
for neutrino decay. For normal mass hierarchy in the absence of NSI the electron neutrino survival probability 
turns out to be small, most of the $\nu_e$'$s$ having been converted through standard oscillations to $\nu_{\mu},
\nu_{\tau}$'$s$. Furthermore it is found that as in the case of the sun, the decay probability is extremely 
small for antineutrinos from neutrino decay to ever be observed. Since in the neutronization phase only 
$\nu_e$'$s$ are produced, no charged current signal is expected to be seen at this initial stage in the 
absence of NSI. On the other hand in the NSI case a sizeable electron neutrino flux may appear at low 
energies in the neutronization phase which may be detected through the charged current 
with improved low energy detectors. At present this remains a challenge but may be reached in the not 
too distant future. For inverse hierarchy all neutrino fluxes are comparable in the whole energy range. 
On the other hand in the thermalization phase neutrinos ($\nu_{\alpha}$) and antineutrinos 
($\bar\nu_{\alpha}$) of all kinds are produced so that the initial state is assumed to consist of 
$\nu_{\alpha}$, $\bar\nu_{\alpha}$'$s$ in equal proportions. Owing to the large number of oscillations, 
the information from the initial state is essentially lost, hence the final probability distributions 
are the same as in neutronization where only $\nu_e$'$s$ are present initially. The only difference is 
in this case the obvious presence of antineutrinos with the same probability distribution as neutrinos 
due to $CPT$ invariance. 
%kinds are produced and the new initial states $\bar\nu_e$, $\nu_{X},\bar\nu_{X}$ (with $X=\mu,\tau$) 
%provide in the NSI case a substantial increase in the appearance probability of $\nu_e$ and $\bar\nu_e$. 
%so that, although their cross section is well below the one for $\bar\nu_e$, a significant $\nu_e$ rate 
%may be observed. 
Finally in section 4 we comment on our results and draw our conclusions.
%the neutronization signal is expected to be small. On the other
%hand, in the NSI case, antineutrinos appear to be abundantly produced from the decay of the neutronization
%$\nu_e$'s in the initial 10 ms burst. This opens the possibility of a clear detection of $\bar\nu_e's$ 
%through (\ref{antinu}) before the thermal neutrino phase. Finally in section 4 we comment on our results 
%and draw our conclusions.
%which was found to explain the absence of the LMA probability upturn in
%the sun. Such an effect in the sun, while being able to explain the absence of
%the upturn of the LMA probability in the intermediate energy range 

\section{The NSI Hamiltonian}

In this section we highlight and discuss the main steps 
of the analysis leading to the Hamiltonian for propagation in dense matter (see also\cite{Pulido:2010ht}).
Its results will be used in section 3 for the evaluation of the probabilities for neutrino survival and 
conversion and the probability for antineutrino production.

Assuming only standard interactions (SI), the potential for electron neutrinos traveling through the sun and supernova
is given by
\be
V=G_F\sqrt{2}N_e\left(1-\frac{N_n}{2N_e}\right)=V_c+V_n
\label{VSI}
\ee
where $N_e$, $N_n$ are the electron and neutron density and $V_c=G_F\sqrt{2}N_e$, $V_n=-G_F/\sqrt{2}N_n$.
For NSI we assume that the $\nu_{\alpha}$ interaction potential on electrons $(\alpha=e,\mu,\tau)$ involves both 
the charged and neutral currents (CC and NC), while on quarks it involves only NC. NSI give rise to possible 
lepton flavour violation. Denoting by $\varepsilon^{fP}_{\alpha \beta}$ the NSI factor that multiplies each 
diagram associated to neutrino propagation in matter we have
\begin{eqnarray}
(v_{\alpha\beta})_{NSI}\!\!\! & = & \!\!\!G_F\sqrt{2}N_e\left[(\varepsilon_{\alpha \beta}^{eP})_{NC}+\left(-\frac{1}{2}+2sin^2\theta_W \right)
(\varepsilon_{\alpha \beta}^{eP})_{NC}+\left(1-\frac{8}{3}sin^2\theta_W+\frac{N_n}{2N_e} \right) 
\varepsilon_{\alpha \beta}^{uP}\right.\nonumber\\
&& + \left.\left(-\frac{1}{2}+\frac{2}{3}sin^2\theta_W-\frac{N_n}{N_e} \right)
\varepsilon_{\alpha \beta}^{dP} \right]
\label{VNSI}
\end{eqnarray}
In the following we will assume $(\varepsilon_{\alpha \beta}^{eP})_{CC}=(\varepsilon_{\alpha \beta}^{eP})_{NC}$. Hence
flavour change may occur without a vacuum mixing angle \cite{Fogli:2008ig,Schwetz:2008er} or a magnetic 
moment \cite{Das:2009kw}, being induced only by the off diagonal entries of this matrix ($\alpha\neq\beta$).
So with SI and NSI the matter Hamiltonian in the flavour basis is the sum of eqs.(\ref{VSI}) and (\ref{VNSI})
\be
{\cal H}_M=V_c \left(\begin{array}{ccc} 1 & 0 & 0\\
0 & 0 & 0\\
0 & 0 & 0\\ \end{array}\right)+
\left(\begin{array}{ccc} (v_{ee})_{NSI} & (v_{e\mu})_{NSI} & (v_{e\tau})_{NSI}\\
(v_{\mu e})_{NSI} & (v_{\mu\mu})_{NSI} & (v_{\mu\tau})_{NSI}\\
(v_{\tau e})_{NSI} & (v_{\tau\mu})_{NSI} & (v_{\tau\tau})_{NSI}\\ \end{array}\right)={\cal H_{SI}}+{\cal H_{NSI}}~.
\label{HNSI}
\ee
As is well known, owing to the large neutrino density, collective effects can play an important role in 
the supernova \cite{Pastor:2002we}, \cite{Duan:2006jv}, \cite{Duan:2006an} \footnote{For a review on 
collective oscillations see \cite{Duan:2010bg}.}, providing an additional contribution to  
the matrix (\ref{HNSI}). They occur up to a few 100 km, whereas MSW oscillations occur 
typically at larger distances so that MSW effects factorize and can be included separately \cite{Dasgupta:2007ws}. 
For this reason the NSI effects, intrinsically associated to MSW, must also be included separately. 
We will perform our calculation starting from a region around 400-500 km where collective oscillations have already
taken place. Hence their net effect to our approach amounts to the well known spectral swap-split of the neutrino 
and antineutrino energy spectra \cite{Duan:2006jv}, \cite{Duan:2006an}, \cite{Dasgupta:2009mg}. However it was
recently shown that matter completely suppresses collective oscillations up to 200 ms \cite{Chakraborty:2011nf} 
after bounce. In view of these results it appears that collective effects can be ignored at early times in a 
supernova. Our results, derived in the next section, are therefore expected to be valid for the whole neutronization 
phase and part of the subsequent thermal phase.
%so that 
%this change in the spectra can be ignored in our analysis provided it is restricted to the initial stages
%of supernova evolution. Our results, derived in the next section where we omit collective effects, are therefore 
%valid for the whole neutronization phase and part of the subsequent thermal phase.

The investigation performed in ref.\cite{Pulido:2010ht} shows that no off diagonal entry in matrix 
(\ref{HNSI}) whether real or imaginary, can change 
the LMA probability, hence the rates and the corresponding SuperKamiokande and SNO electron spectra. In fact 
only imaginary diagonal couplings lead to a change in $P_{LMA}$. As discussed in ref. \cite{Pulido:2010ht}, 
this implies the instability of neutrinos in matter and their decay into antineutrinos of all species along
with majorons. Requiring the convenient change in $P_{LMA}$, namely the one that leads to a flat electron 
spectrum in SuperKamiokande and SNO, thus allowing for imaginary diagonal couplings, it was shown that the 
simplest choice of parameters is  
\begin{equation}
{\cal H_{NSI}}=G_F\sqrt{2}N_e\!\!\left[\!\!\left(\begin{array}{ccc}
\!\!\frac{i}{2}\varepsilon (x_e+1) &  &\\ & -i\varepsilon x_e &\\ & & \frac{i}{2}\varepsilon x_e \!\!\end{array} \right)
\!\!+\!x_u\!\!\left(\begin{array}{ccc} \!\!\frac{i}{2}\varepsilon & &\\ & -i\varepsilon &\\
&  & \frac{i}{2}\varepsilon\!\! \end{array} \right)\!\!+
\!x_d\!\left(\begin{array}{ccc} \!\!-\frac{i}{2}\varepsilon &  &\\ & i\varepsilon &\\ &  &
-\frac{i}{2}\varepsilon\!\! \end{array} \right)\!\!\right]
\label{HNSIf}
\end{equation}
with vanishing off diagonal entries and $\varepsilon=3.5\times 10^{-4}$. Here
\be
x_e=-\frac{1}{2}+2sin^2\theta_W,~x_u=1-\frac{8}{3}sin^2\theta_W+
\frac{N_n}{2N_e},~x_d=-\frac{1}{2}+\frac{2}{3}sin^2\theta_W-\frac{N_n}{N_e}~.
\label{x}
\ee
The three matrices in the right hand side of eq. (\ref{HNSIf}) relate to the neutrino interaction with 
electrons, u-quarks and d-quarks respectively. Each diagonal entry refers to the $\nu_e$, $\nu_{\mu}$, 
$\nu_{\tau}$ contribution to its own interaction, hence its decay. So for instance, for $\nu_e$ the NSI (decay)
%the interaction with electrons, u-quarks 
%and d-quarks to the propagation (hence the decay) of $\nu_{\alpha}$. So for instance, for $\nu_e$ the NSI (decay) 
potential is 
\be
(v_{ee})_{NSI}=\frac{i}{2}\varepsilon G_F\sqrt{2}N_e(x_e+1+x_u-x_d)~.
\label{vee}
\ee
In this way we obtain the three interaction (decay) potentials
\be
(v_{ee})_{NSI} = iG_F\sqrt{2}(3.5\times 10^{-4})N_e\left(1-\frac{2}{3}sin^2\theta_W+\frac{3N_n}{4N_e}\right)
\label{ve}
\ee
\be
(v_{\mu\mu})_{NSI} = iG_F\sqrt{2}(3.5\times 10^{-4})N_e\left(-1+\frac{4}{3}sin^2\theta_W-\frac{3N_n}{2N_e}\right)
\label{vmu}
\ee
\be
(v_{\tau\tau})_{NSI} = \frac{i}{2}G_F\sqrt{2}(3.5\times 10^{-4})N_e\left(1-\frac{4}{3}sin^2\theta_W+\frac{3N_n}{2N_e}\right)~.
\label{vtau}
\ee
The full Hamiltonian including the vacuum part and referred to the mass basis is now
\be
{\cal H}=\left(\begin{array}{ccc} 0 & 0 & 0\\
0 & \frac{\Delta m^2_{21}}{2E_0} &  0\\
0 &       0         &  \frac{\Delta m^2_{31}}{2E_0}\\ \end{array} \right)+ U^{\dagger}
{\cal H}_M U~
\label{H}
\ee
with ${\cal H}_M$ given by (\ref{HNSI}) with the replacements (\ref{ve}), (\ref{vmu}), (\ref{vtau}) and arbitrary
off diagonal entries. In eq. (\ref{H}) $\Delta m^2_{21}$, $\Delta m^2_{31}$ are the solar and atmospheric mass 
differences, $U$ is the PMNS matrix \cite{Maki:1962mu} defined with the standard 
parameterisation \cite{Amsler:2008zzb} and we use the central value for $sin\theta_{13}=0.13$ from ref. 
\cite{Fogli:2008ig}. The negative imaginary parts of the eigenvalues of (\ref{H}) are the mass eigenstate decay 
rates $\Gamma_i$ which will be used in the next section for the evaluation of the probabilities. As for the
supernova parameters, numerical simulations \cite{Fischer:2009af} yield the electron number density and 
supernova density profiles which in our period of interest (the initial 200 ms) are well approximated by
\be
Y_e=\frac{1}{3}-0.04~log~\frac{\rho}{10^{12}g~cm^{-3}}
\ee
\be
\rho=\rho_{0}\left(\frac{10km}{r}\right)^3 g~cm^{-3}
\ee
for $r>10~km$ with $\rho_{0}\sim 10^{14}g~cm^{-3}$ and which will be used in the next section.

\section{Rates, couplings and probabilities}

\subsection{Rates and neutrino majoron couplings}

\begin{figure}
\centering
\hspace*{-1.5cm}
\includegraphics[height=200mm,keepaspectratio=true,angle=0]{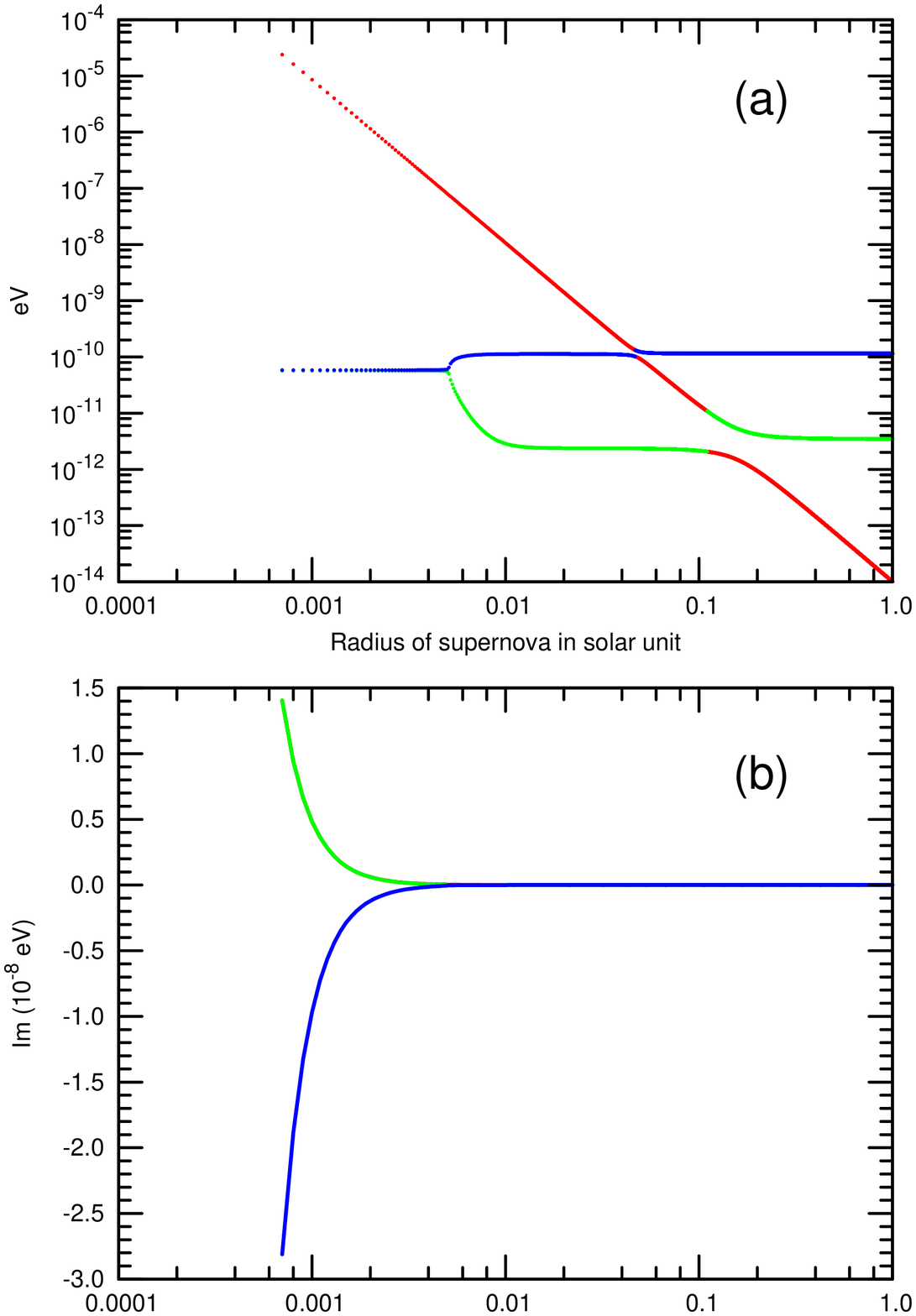}
\caption{ \it The real (a) and imaginary parts (b) of the neutrino mass matter eigenvalues for
neutrino energy $E_0=11MeV$. The two resonances ('atmospheric' and 'solar') are clearly
visible in panel (a) and the two positive imaginary parts are superimposed in panel (b).}
\label{fig1}
\end{figure}

The $\nu_e$ flux from neutronization is in fact a linear combination of the three mass eigenstates $\nu_i$ 
displayed in fig.\ref{fig1}({\it a}) for neutrino energy $E_0=11MeV$ and normal hierarchy. Our first purpose 
in this section is to evaluate the decay rate of the NSI process \cite{Berezhiani:1987gf,Berezhiani:1989za,
Kachelriess:2000qc,Tomas:2001dh,Farzan:2002wx,Lessa:2007up}
\be
\nu_{i}\rightarrow \bar\nu_{j}+\chi
\ee
where $\chi$ denotes the majoron. This rate satisfies
\cite{Pulido:2010ht}, \cite{Farzan:2002wx}
\be
\frac{\partial \Gamma_{i}}{\partial E_f}=\sum_{j=1}^{3}\frac{|g_{ij}|^2}{8\pi}\frac{E_0-E_f}{{E_0}^2}
|v_{i}(r)-\overline v_{j}(r)|_{NSI}F(r,E_0)
\label{eq3}
\ee
where $g_{ij}$ are the neutrino majoron couplings \cite{Farzan:2002wx},\cite{Lessa:2007up} and the interaction 
potentials satisfy in the mass basis $v_i=-\overline v_i$. Here $E_0$ is the initial neutrino energy, $E_f$ is 
the antineutrino energy which in a first approximation we assume to take values in the interval $(0,E_0)$, 
since the energy $E_0$ is shared by the final neutrino and the majoron. The quantity $F(E_0,r)$ is
the Fermi factor \cite{Farzan:2002wx}
\be
F(E_0,r)=\left(1-\frac{1}{e^{(E_{0}-\mu)/T}+1}\right)~,
\label{FF}
\ee
which reflects the fact that inside the supernova some of the states have already been occupied by neutrinos.
In the inner core ($R_{inner}\simeq 10~km$) the chemical
potential for $\nu_e$ $(\mu_{\nu_e}$) is around 200 MeV and the temperature $T=35~MeV$. In the outer core
($R_{inner}\simeq 15~km$) the temperature drops abruptly $T=2~MeV$, the density falls from $5\times 10^{14}g~cm^{-3}$
to $5\times 10^{13}g~cm^{-3}$ and we may set $\mu_{\nu_e}=0$. Hence in the following we will omit the factor
$F(E_0,r)$.

The three decay rates $\Gamma_i$ are the imaginary parts of the NSI Hamiltonian eigenvalues. They are 
represented in fig.\ref{fig1}({\it b}) for $E_0=11MeV$. As in the solar case, only $\Gamma_3$ is
negative, so only the state $\nu_3$ is unstable, allowing for the decay into either $\bar\nu_1$,
$\bar\nu_2$ or $\bar\nu_3$, thus generating antineutrinos of the three flavours. One could obtain
an alternative expression for $\Gamma_{i}(r,E_0)$ through the integration of equation (\ref{eq3}) 
over $E_f$
\be
\Gamma_i=\int_0^{E_0}\frac{\partial \Gamma_{i}}{\partial E_f}dE_f=\sum_{j=1}^{3} \frac{|g_{ij}|^2}{16\pi}
|v_{i}(r)-\bar v_{j}(r)|~.
\label{intgamma}
\ee
The vanishing lower limit used in this integration is as referred to above only an approximation, since the 
majoron obtains a tiny effective mass in matter $m_{eff}^2\sim |g|^2N_{\nu}/q$ \cite{Farzan:2002wx}. In this 
way a slight dependence of the rate $\Gamma_{i}$ on $E_0$ arises which is consistent with the numerical 
evaluation of the imaginary part of the NSI Hamiltonian eigenvalues. 

The parameters $\varepsilon_{\alpha\beta}$ were fixed earlier (see section 2) by the fittings to the solar 
neutrino data \cite{Pulido:2010ht} and so the
rates $\Gamma_{i}(r,E_0)$, numerically evaluated as the imaginary parts of the Hamiltonian eigenvalues, are 
also fixed. Moreover the values of $g_{ij}$ to which the rates correspond are so far unknown and only upper
bounds exist in the literature \cite{Lessa:2007up}. In the following we will use the strictest one quoted, 
namely 
\be
\sum_{\alpha}|g_{e\alpha}|^2<5.5\times 10^{-6}, 
\label{bound}
\ee
conveniently expressed in the mass basis. Quantities $\Gamma_i$ will now be used in the evaluation of the
probabilities.
%We are therefore now in a position to 
%provide an estimation of the sum of the neutrino majoron couplings $g_{ij}$ by fitting expression 
%(\ref{intgamma}) to the imaginary NSI Hamiltonian eigenvalues given in fig.1.

\subsection{Probability densities for neutrino survival, decay and antineutrino appearance}

We denote by $P_{\nu_{i}}(r,E_0)$ the $\nu_i$ survival probability in the mass
basis for neutrino energy $E_0$ at a distance $r$ from the star centre, obtained from integration of the 
Schr\"{o}dinger equation with the Hamiltonian (\ref{H}), and by $\phi_{\nu_e}(E_0)$ the initial normalized 
neutrino spectral flux \cite{Thompson:2002mw}
$$\phi_{\nu_e}(E_0)=\frac{1}{\Phi(E_0)}\frac{\partial \Phi}{\partial E_0}~.$$
Given these definitions the quantity
\be
\phi_{\nu_i}(r,E_0)\!=\!P_{\nu_{i}}(r,E_0)\phi_{\nu_e}(E_0)
\label{eqprob}
\ee
is the normalized spectral flux of $\nu_i$ mass eigenstates with energy $E_0$ that remain in the beam after traveling a 
distance $r$. Hence  
\be
\frac{\partial P_{\nu_{i}^m}(E_0)}{\partial r}=
\int_{0}^{E_{0}}\phi_{\nu_i}(r,E_0)(1-e^{\Gamma_{i}(r,E_0)r})~
\frac{\partial \Gamma_{i}(r,E_0,E_f)}{\partial E_f}~dE_f
\label{eq1}
\ee
is the probability per unit star radius for this mass eigenstate to have disappeared from the flux. %\footnote{

Replacing now 
(\ref{eq3}) in (\ref{eq1}) one obtains the probability for the $\nu_i$ mass eigenstate disappearance 
\be
P_{\nu_{i}^m}(E_0)\!=\sum_{j=1}^{3}\!\frac{|g_{ij}|^2}{8\pi}\!\!\int_{R_i}^{R_S}\!\!|v_{i}(r) - \overline v_{j}(r)|_{NSI}
\phi_{\nu_i}(r,E_0)(1-e^{\Gamma_{i}(r,E_0)r})\!\!\int_{0}^{E_0}\!\frac{E_0\!-\!E_f}{{E_0}^2}dE_f dr
\label{eq5}
\ee
Quantities $R_i$ and $R_S$ denote the neutrino sphere and the star radii respectively.
We note that in eq.(\ref{eq5}) we have the simple sum of probabilities, since each eigenstate $\nu_i$, as it 
decays, can give rise to just one antineutrino flavour which is a linear combination of the three mass 
eigenstates $\nu_j$. Performing the integration in $E_f$, (\ref{eq5}) can be simplified to
\be
P_{\nu_{i}^m}(E_0)\!=\sum_{j=1}^{3}\!\frac{|g_{ij}|^2}{16\pi}\!\!\int_{R_i}^{R_S}\!\!|v_{i}(r) - \overline v_{j}(r)|_{NSI}
~\phi_{\nu_i}(r,E_0)(1-e^{\Gamma_{i}(r,E_0)r})~dr~.
\label{eq6}
\ee
In other words, given a flux of $\nu_i$' s with an energy in the interval $[E_0,E_{0} + dE_{0}]$, the quantity
$P_{\nu_{i}^m}(E_0)dE_0$ is the fraction of these neutrinos which has decayed into antineutrinos with an
energy in the interval $(0,E_0 + dE_0)$ after traversing the star. The fraction of $\nu_i$' s that remains 
in the beam after leaving the star is $P_{\nu_{i}}(R_s,E_0)dE_0$ with $P_{\nu_{i}}(r,E_0)$ as defined in the 
beginning of this section (see (\ref{eqprob})).

\begin{figure}
\centering
\hspace*{-1.5cm}
\includegraphics[height=200mm,keepaspectratio=true,angle=0]{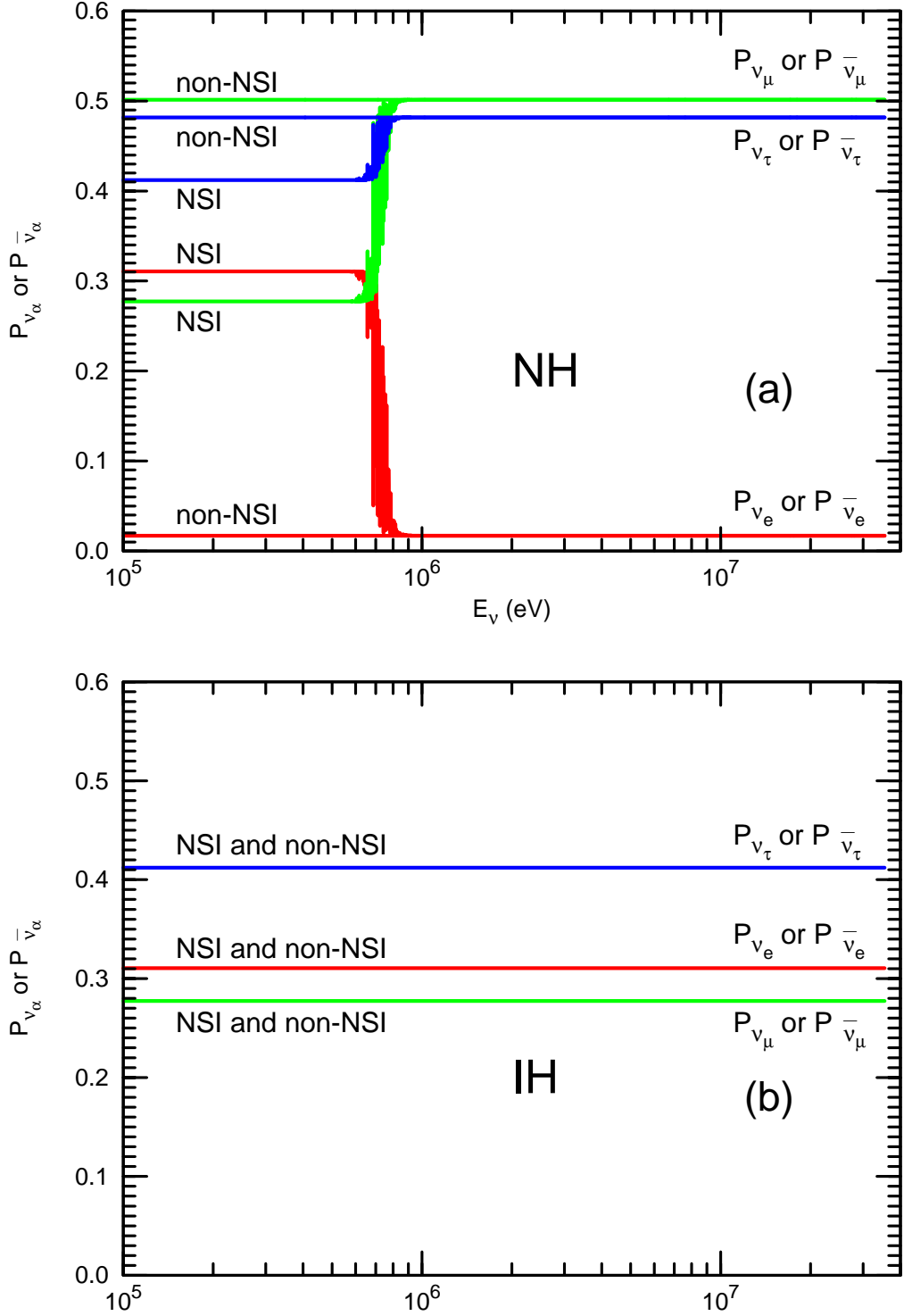}
\caption{ \it The appearance probabilities $P_{\nu_e},~P_{\bar\nu_e},~P_{\nu_{\mu}},
~P_{\bar\nu_{\mu}},~P_{\nu_{\tau}},~P_{\bar\nu_{\tau}}$
with and without NSI ((a) normal hierarchy, (b) inverse hierarchy). For normal hierarchy
and energy below 0.9 MeV the NSI probability merges with its non-NSI value.}
%For the initial neutronization phase only neutrinos are expected. 
%Antineutrinos are expected in the thermalization phase along with the former. Their probability profiles 
%are the same as the neutrino ones.}
\label{fig2}
\end{figure}

As a reminder we note that the normalization of $P_{\nu_{i}}$ follows from
%As we have seen, the integration of the equation of motion proceeds in the mass basis, so that the normalization
%condition is forced on $P_{\nu_{i}}$. We have
\be
P_{\nu_i}=|<\nu_i|\nu_e>|^2=|\nu_i u_{ej}^{*}\nu_j|^2=u_{ej}^{*}u_{ek}\nu_{j}\nu_{k}\nu_{i}\nu_{i}=
u_{ej}^{*}u_{ek}\delta_{ij}\delta_{ik}=u_{ei}^{*}u_{ei}
\label{probmass}
\ee
(no sum over $i$) with the orthogonality of $U^{PMNS}$ ensuring $\sum_{i}P_{\nu_i}=1$. Moreover the integration of
the equation of motion proceeds in the mass basis, so that the normalization condition is forced on $P_{\nu_{i}}$.
The normalization of $P_{\nu_{\alpha}}$ also follows in a similar way
\begin{eqnarray}
P_{\nu_{\alpha}}&=&|<\nu_{\alpha}|\nu_e>|^2=|u_{\alpha i}^{*}u_{ej}\nu_{i}\nu_{j}|^2 = u_{\alpha i}^{*}u_{ej}^{*}
u_{\alpha k}u_{em}\nu_{i}\nu_{j}\nu_{k}\nu_{m}  \nonumber \\ & = & u_{\alpha i}^{*}u_{ej}^{*}u_{\alpha j}u_{ei}=
\delta_{\alpha e}u_{\alpha i}^{*}u_{ei}
\label{probflavour}
\end{eqnarray}
(no sum over $\alpha$) with the orthogonality of $U^{PMNS}$ implying $\sum_{\alpha}P_{\nu_{\alpha}}=1$. In
(\ref{probmass}) and (\ref{probflavour}) we neglected the plane wave propagation phases. Finally the flavour 
probability is obtained from
\be
P_{\nu_{\alpha}}=|u_{\alpha i} <\nu_i|\nu_e>|^2
\ee
and is shown in fig.\ref{fig2} for standard ($\varepsilon=0$) and non-standard interactions ($\varepsilon=3.5\times 10^{-4}$).

For the sake of the following discussion we recall that from supernova theory only electron neutrinos are
produced in the initial neutronization phase. From fig.\ref{fig2} with normal hierarchy (panel (a)) it is seen that, 
in the absence of NSI, electron neutrinos can hardly be detected, as the survival probability is around $0.02$ and 
practically constant with energy. Further, $\nu_{\mu}$'$s$ and $\nu_{\tau}$'$s$ produced from oscillations, 
although having a much larger production probability ($\sim 0.5$), can only be detected through neutral current 
reactions, so there will be no way to distinguish them at present from other neutrinos or antineutrinos. 
%Hence 
%in the absence of NSI we are left with the possibility of tracing only electron antineutrinos from the thermal phase.
%The remainder ($\nu_X,\bar\nu_X$ with $X=\mu,\tau$) although possible to detect through the neutral current, will be 
%indistinguishable from each other. 
The major difference in the NSI case for normal hierarchy as seen from fig.\ref{fig2}a is the appearance of a 
visible flux of electron neutrinos in the low energy region ($E_0\lesssim 0.8-0.9MeV$). This can be pinpointed in 
the neutronization phase through the charged current, provided enough experimental capability is developed in the
future to reduce 
the low energy threshold. On the other hand for inverse hierarchy (fig.\ref{fig2}b) no difference appears between NSI
and non-NSI: the sizable value of the $\nu_e$ probability ($\sim 0.31$) is the same at low energy as in the normal
hierarchy case and remains constant with energy, so $\nu_e$ detection is experimentally accessible. However the
NSI and non-NSI probability values are the same for each neutrino species and also remain in each case constant with 
energy. So for inverse hierarchy NSI and non-NSI appear indistinguishable.
%Electron antineutrinos will accompany them in the thermal phase.

\begin{figure}
\centering
\hspace*{-1.5cm}
\includegraphics[height=120mm,keepaspectratio=true,angle=0]{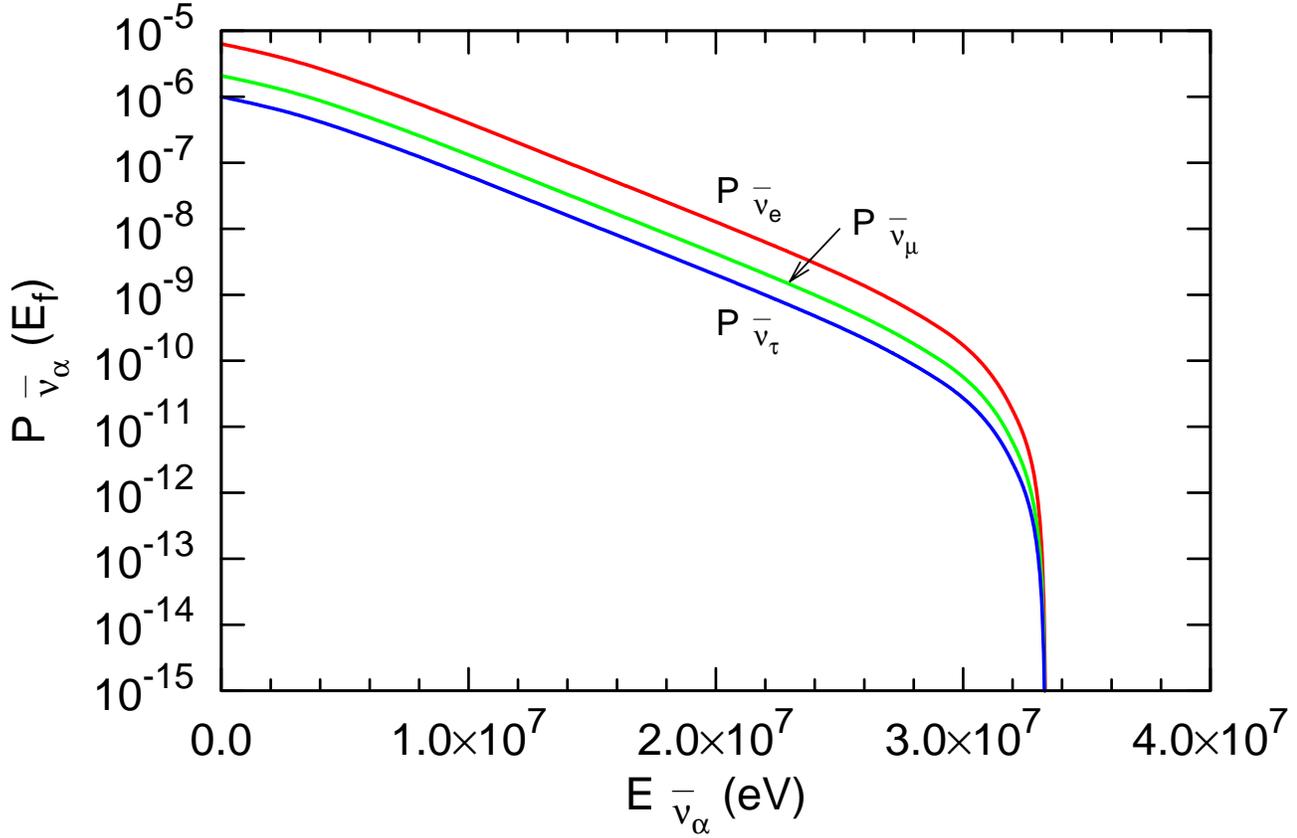}
\caption{ \it The antineutrino appearance probabilities from neutrino decay in matter through NSI.
The upper bound (\ref{bound}) (the strictest) was used as a common value for the neutrino majoron couplings.
Antineutrinos from neutrino decay in supernova matter cannot be detected even if the more conservative 
bound (\ref{bound1}) is used.}
\label{fig3}
\end{figure}

We next investigate the possibility for electron antineutrinos to be detected in the initial neutronization phase. As
previously referred, only electron neutrinos are produced at this stage in a 10 ms pulse. Their decay in matter into 
electron antineutrinos in a significant number would provide a clear signature of NSI in this initial pulse. 
Antineutrinos with energy $E_f$ are produced from electron neutrino decay with energy $E_0$ in the interval 
$(E_f,E_{0_{max}}]$. Their appearance probability density is in the mass basis
\be
P_{\bar\nu_{j}}(E_f)\!={\bigcup}_{i=1}^{3}\!\frac{|g_{ij}|^2}{8\pi}\!\!\int_{R_i}^{R_S}\!\!|v_{i}(r)-\overline v_{j}(r)|_{NSI}\!
\!\int_{E_{f}}^{E_{0_{max}}}\!\!\!\phi_{\nu_{i}}(r,\!E_0)(1-e^{\Gamma_{i}(r,E_0)r})\frac{E_0\!-\!E_f}{{E_0}^2}dE_0dr.
\label{antinu}
\ee
Contrary to eq.(\ref{eq6}), where each neutrino could decay into one {\it single} antineutrino, each antineutrino
can be now produced from more than one neutrino, hence the reason to consider the union of events in eq.(\ref{antinu})
\footnote{The probability for the union of three independent events ($A_1,A_2,A_3$) is given by the well known rule
$P(A)=P(A_1)+P(A_2)+P(A_3)-P(A_1)P(A_2)-P(A_1)P(A_3)-P(A_2)P(A_3)+P(A_1)P(A_2)P(A_3)$.}.
Assuming CPT invariance the flavour probability for $\bar\nu_e$ is given by
\be
P_{\bar\nu_{e}}=|u_{e i}|^2 P_{\bar\nu_{i}}~.
\label{antiprob}
\ee
Using the bound (\ref{bound}) as a common value for all neutrino majoron couplings involved, we obtain the result 
displayed in fig.\ref{fig3} for $P_{\bar\nu_{e}}$ which is obviously too small for ${\bar\nu_{e}}$'$s$ to be observed.
If one considers the $\nu_e$ decay into $\bar\nu_{\tau}$ one may relax (\ref{bound}) and use instead \cite{Lessa:2007up}
\be
\sum_{\alpha}|g_{\tau\alpha}|^2<5.5\times 10^{-2}
\label{bound1}
\ee
as a common value for all couplings. In this case the uppermost value of the probability is raised by a factor  
$O(10^4)$ which is still too small for the effect to be observed and moreover $\bar\nu_e$'$s$, if produced 
from $\nu_{\tau}$ decay, would appear as a higher order effect. On the other hand, as pointed above, the $\nu_e$  
probability from oscillations in the absence of NSI is seen to be rather small in normal hierarchy (fig.\ref{fig2}a). 
The prospects for its detection and hence a charged current signal in the neutronization phase will crucially depend 
on the detector size and supernova distance. With NSI, it may become clearer through $\nu_e d\rightarrow 
ppe^{-}$ or an increased $\nu e^-\rightarrow \nu e^-$ scattering event rate, however at low energy 
($E_0\lesssim 0.8-0.9MeV$), which is still an experimental challenge. For inverse hierarchy, as also pointed above,
the situation is much different: the $\nu_e$ signal appears louder and clearer (see fig.\ref{fig2}b).
%Hence in the absence of NSI, from the  
%%$\bar\nu_{\mu}$ and $\bar\nu_{\mu}$ through NSI decay of $\nu_e$
%%As displayed in fig.\ref{fig3} this is extremely small and cannot be observed. 
%%Hence in the absence of NSI, recalling 
%figs.\ref{fig2} and \ref{fig3}  no charged current signal is 
%expected in the initial neutronization phase. With NSI it may appear at low energy through $\nu_e p\rightarrow 
%ppe^{-}$ or an increased $\nu e^-\rightarrow \nu e^-$ scattering event rate.

In the subsequent thermalization phase $\bar\nu_e$'$s$ and $\nu_X,\bar\nu_X$'$s$ with $X=\mu,\tau$ are produced 
initially along with $\nu_e$'$s$. All kinds of neutrinos and antineutrinos will arise from
these through NSI. The appearance probability for antineutrinos from neutrino decay may increase substantially,
since the bound (\ref{bound1}) must now obviously be taken into account instead of (\ref{bound}). Again this 
is not expected to be enough for the effect to be observable (see fig.\ref{fig3}) even for $\bar\nu_e$ 
appearance, despite the higher rate involved in its detection. As for the other probabilities from oscillations 
and NSI, they must be evaluated from the union of events as in (\ref{antinu}), since each final neutrino or 
antineutrino is produced simultaneously from a number of different initial ones. Evaluating these probabilities 
taking into account their normalization, it turns out that the contribution from the extra initial neutrinos
and antineutrinos does not change their value nor energy distribution relative to the neutronization
case when initially only $\nu_e$'$s$ are present. In fact the information from the initial state is lost
due to the exceedingly large number of oscillations that the neutrinos undergo during propagation: only the 
propagation physics which depends on the interaction potentials is relevant here. Thus the same fig.\ref{fig2}
applies for thermalization both for normal and inverse hierarchies. The only notorious characteristic to tell 
NSI from non-NSI is, as in neutronization and normal hierarchy, 
the comparatively large probability for $\nu_e$, $\bar\nu_e$ at low energy ($E_0\lesssim 0.8-0.9MeV$) with the 
same energy profile. Such low energy raises however an experimental challenge for detection. For inverse 
hierarchy $\nu_e$'$s$ and $\bar\nu_e$'$s$ remain equally abundant both for $E_0\lesssim 0.8-0.9MeV$ and larger, 
therefore their detection is acessible, although NSI and non-NSI are indistinguishable in this case (see 
fig.\ref{fig2}b). In particular $\bar\nu_e$'$s$ provide a clear signal through the reaction $\bar\nu_e p
\rightarrow ne^+$ whose cross section is $O(10^2)$ larger than for scattering with electrons.
Since thermalization is a much longer process than neutronization ($>$10 s), a larger accumulation of events 
is possible in this phase.

For the detection and measurement of the $\nu_{\mu}$, $\bar\nu_{\mu}$, $\nu_{\tau}$, $\bar\nu_{\tau}$ individual 
energy spectra which can only be traced via neutral currents, an interesting proposal was presented some time 
ago \cite{Beacom:2002hs} and recently revived \cite{Dasgupta:2011wg}. It is based on the $\nu~p\rightarrow 
\nu~p$ scattering reaction which can be observed in scintillator detectors (e.g. Borexino, SNO+, KamLAND) 
through their adequate preparation. This is a neutral current process with a cross section about $O(10^2)$ 
larger than neutrino electron scattering at supernova neutrino energies. For the NSI scenario expound in the 
present paper this technique will be particularly useful, since it appears to be possible to clearly distinguish 
between normal and inverted hierarchies. In fact it suffices to note that for normal hierarchy the above 
mentioned neutrinos arrive copiously on Earth in comparison with the more rare $\nu_e$$'$s and $\bar\nu_e$$'$s 
whereas for inverse hierarchy all species arrive in comparable numbers (see fig.\ref{fig2}a, b).

\section{Summary and conclusions}

We have extended to the supernova the previously developed model for neutrino NSI in the sun introduced earlier 
to remove the tension between the LMA predictions and the experimental signatures of solar neutrinos, especially 
the absence of an upturn in the SuperKamiokande event rate. Improving the data fittings in the solar case 
implies neutrino decay in dense matter into antineutrino and a majoron, hence the motivation to investigate
the consequences of the model for the supernova. 

In the present paper we found however that, although the matter density in supernova is much larger than 
in the sun, the extension of neutrino trajectory in the very high density medium is too short to imply
a significant neutrino decay into antineutrino and a majoron and the corresponding appearance probability is 
insignificant. In the initial and short neutronization phase (10 ms) where only neutrinos are produced 
through the reaction (\ref{neutron}) no antineutrinos are expected experimentally whether or not NSI applies. 
The important NSI trace is the $\nu_e$ appearance probability which increases from 0.02 to 0.31 at low 
energies ($E_0\lesssim 0.8-0.9MeV$) for normal hierarchy, while for inverse hierarchy NSI and non-NSI
cannot be distinguished. In this case the $\nu_e$ probability remains at 0.31 regardless of the energy.
In the neutronization (deleptonization) phase $\nu_e$'$s$ are the only states that can induce charged 
current interactions, so they can be singled out either through an increased $\nu e^{-} \rightarrow 
\nu e^{-}$ scattering event rate or $\nu_e d\rightarrow ppe^{-}$. The remainder ($\nu_{\mu}$'$s$, 
$\nu_{\tau}$'$s$) inducing only neutral currents, cannot be distinguished from each other nor from 
$\nu_{e}$'$s$. Detecting these $\nu_e$'$s$ remains however an experimental challenge at present in
normal hierarchy, but not so for inverse hierarchy, as they appear more copiously at higher energies. 
%The important and observable NSI trace will be 
%the increase in the $\nu_e$ probability relative to non-NSI. 
%In the initial and short (10 ms) neutronization phase the $\nu_e$ flux is moderately 
%increased with respect to its non-NSI value: whereas in the absence of NSI the probability is practically 
%energy independent and small (0.02), for NSI it increases to 0.31 at the low energy end ($E\lesssim 0.8-0.9MeV$). 
In the subsequent and longer thermalization phase, the extra neutrino and antineutrino states ($\bar\nu_e$ and $\nu_X, 
\nu_{\overline X}$ with $X=\mu,\tau$) that are produced through processes (2)-(5) cannot change the appearance 
probabilities relative to the neutronization phase. Hence detecting these $\nu_e$'$s$ and $\bar\nu_e$'$s$ is, 
again, an experimental
%due to NSI is from 0.05 to 0.66 which means that $\nu_e$
%and $\bar\nu_e$'$s$ are copiously produced, however at low energy. Detecting them is, again, an experimental
challenge in normal hierarchy: as for the neutronization phase $\nu_e$'$s$ can be detected through a major event 
rate increase originated from the charged current in $\nu e^{-}\rightarrow \nu e^{-}$ scattering or through the 
reaction $\nu_e d 
\rightarrow p p e^{-}$, while $\bar\nu_e$'$s$ through the clear signal $\bar\nu_e p \rightarrow n e^{+}$. The 
remaining neutrinos and antineutrinos can only be detected through the neutral current and so cannot be distinguished 
from $\nu_e$'$s$ and $\bar\nu_e$'$s$, unless the interesting technique proposed in \cite{Beacom:2002hs}, 
\cite{Dasgupta:2011wg} is developed. If and when this advancement succeeds, it may be possible within the present 
scenario to tell normal from inverse hierarchy. As regards collective oscillations, their effect in our analysis amounts 
to the modification of the neutrino and antineutrino spectral fluxes. Collective effects are however probably suppressed 
up to 0.2 seconds after bounce. The results obtained are therefore applicable to the neutronization (deleptonization) 
phase and part of the subsequent thermal phase.

To summarize, in the presence of NSI we expect a sizable flux of $\nu_e$'$s$ and $\bar\nu_e$'$s$ at all energies 
for inverse hierarchy and at low energy ($E_0\lesssim 0.8-0.9MeV$) for normal hierarchy. These fluxes are the same 
as for non-NSI. The clear distinction between NSI and non-NSI is possible only for normal hierarchy at low energy
with a more intense flux of $\nu_e$ and $\bar\nu_e$, whose detection is at present an experimental challenge.
Hence in the absence of NSI the chances for observation in normal hierarchy of a charged current signal do not 
appear much favourable at present, but they will of course mainly depend on the detector 
size and supernova distance. Other neutrinos and antineutrinos are in contrast abundantly 
present, however they can only be detected through the neutral current. As in
the case of the sun the antineutrino appearance probability from NSI neutrino decay is in all cases too small for 
antineutrinos to be detected from this origin.

%section*{Acknowledgments}
%c{\partial \Gamma_{i}}{\partial E_f}=\sum_{j=1}^{3}\frac{|g_{ij}|^2}{8\pi}\frac{E_0-E_f}{{E_0}^2}
%{\em We are grateful to Marco Picariello, Lu\'{\i}s Lavoura and Sergio 
%Palomares-Ruiz for useful discussions.
%C. R. Das gratefully acknowledges a scholarship from Funda\c{c}\~{a}o para
%a Ci\^{e}ncia e a Tecnologia (FCT, Portugal) ref. SFRH/BPD/41091/2007. This work was partially 
%supported by the Marie Curie RTN MRTN-CT-2006-035505 and by Funda\c{c}\~{a}o para
%a Ci\^{e}ncia e a Tecnologia through the projects
%CERN/FP/ 109305/2009,  PTDC/FIS/098188/2008
%and CFTP-FCT UNIT 777  which are partially funded through POCTI
%(FEDER).} 

%\begin{figure}
%\centering
%\hspace*{-1.5cm}
%%\includegraphics[height=120mm,keepaspectratio=true,angle=0]{fig_1_c.eps}
%\includegraphics[height=120mm,keepaspectratio=true,angle=0]{prob_t_c.eps}
%\caption{ \it The neutrino and antineutrino appearance probabilities in the thermalization phase. 
%The differences from fig.\ref{fig2} are originated from the fact that in addition to $\nu_e$ all possible 
%neutrinos and antineutrinos are now initial states. Also notice that $P_{\nu_{\alpha}}=P_{\bar\nu_{\alpha}}$ 
%for $\alpha=e,\mu,\tau$.}
%\label{fig4}
%\end{figure}

\end{document}